# Nonparametric Test for Volatility in Clustered Multiple Time Series

***Erniel B. Barrios***
Monash University
Correspondence: erniel.barrios@monash.edu
***Paolo Victor T. Redondo***
King Abdullah University of Science and Technology

**Abstract**

Contagion arising from clustering of multiple time series like those in the stock market indicators can further complicate the nature of volatility, rendering a parametric test (relying on asymptotic distribution) to suffer from issues on size and power. We propose a test on volatility based on the bootstrap method for multiple time series, intended to account for possible presence of contagion effect. While the test is fairly robust to distributional assumptions, it depends on the nature of volatility. The test is correctly sized even in cases where the time series are almost nonstationary (i.e., autocorrelation coefficient $\approx$ 1). The test is also powerful specially when the time series are stationary in mean and that volatility are contained only in fewer clusters. We illustrate the method in global stock prices data.

# 1. Introduction

With the continuous expansion of storage spaces for digital data, accumulation of multiple time series has become synonymous with engagement of various stakeholders involved in a particular phenomenon. Take as an example stock prices, due to increasing storage space, real-time prices can now be recorded, including short-run spikes that could be triggered by random shocks, e.g., news on replacement of the management of a company. This may not be an issue if stock prices are recorded on lower frequencies (e.g., weekly or monthly). However, many information could easily be lost due to aggregation, hence, prices are recorded per minute or even in seconds. This could trigger high frequency time series data to manifest such shocks as conditional heteroskedasticity (volatility). Furthermore, individual securities behave independently, but contagion within a sector, or a market, a region, or even globally can force a group or all securities to exhibit similar patterns in price movements.

Modeling procedures are available for multiple and multivariate time series data, but these are anchored on distributional assumptions about the random shocks and are greatly affected by irregularities or stylized facts like volatility. Volatility causes perturbation in the dynamic behavior of the process, resulting to more complex data generating process causing difficulty in estimation and produces chaotic forecasts. Volatility has the potential to divert forecasts away from the direction of the time series even after the effect of localized perturbation vanished.

Knowledge on whether volatility is present or not in the time series offers an opportunity to better understand the dynamic behavior of the data, thus facilitating modeling and forecasting. [1] proposed a robust estimation procedure for time series data that exhibit structural change. Furthermore, [2] proposed a test for volatility in a time series data. Predictive ability of estimated models during tranquil period can be enhanced resulting from robust estimation of the model, noted [1]. The aim of this paper is to develop a nonparametric test of volatility in a possibly clustered multiple time series data.

The paper is organized as follows: Section 2 summarizes previous literature on multiple time series and volatility; Section 3 presents the estimation algorithm and the proposed test for volatility in clustered time series; Section 4 discusses results of simulation studies to illustrate the size and power of the proposed test; Section 5 presents the application of the test to actual data; Section 6 summarizes the conclusions.

## 2. Multiple Time Series and Volatility

[3] considered multiple time series data as panel data whose common autoregressive parameter and random effect of individual time series were estimated using generalized method of moments (GMM). Although the method usually fails to converge when the length of time series is larger than the number of time series in the panel, [4] proposed an estimation procedure that incorporates maximum likelihood estimation (MLE) and best linear unbiased predictors (BLUP) into the backfitting algorithm. Simulation studies exhibited the advantages of the hybrid method over Arellano-Bond's GMM estimator in terms of predictive ability specially when the length of the time series is greater than the number of time series in the panel. [4] however noted that the advantages of the method are affected by the variance of the error term (possibly by heteroskedasticity), and to address this problem, [5] proposed an estimation procedure that is robust to conditional heteroskedasticity in the multiple time series. Even in the presence of volatility, [5] observed improvement in parameter estimates as well as the predictive ability of the fitted model. This is however still affected by localized non-stationarity induced by the block bootstrap method even if there is really no global heteroskedasticity in the time series. [4] and [5] are the basis for the postulated multiple time series presented in Section 3.

Volatility in time series has been typically assessed by incorporating models for conditional heteroscedasticity into the model structure, e.g., autoregressive conditional heteroskedastic (ARCH) model [6], generalized autoregressive conditional heteroskedastic (GARCH), [7]. Other volatility models which address issues regarding ARCH- and GARCH-like violation of the non-negative constraints for the variances are also proposed, e.g., exponential generalized autoregressive conditional heteroskedastic model (EGARCH), [8]. But these more general models also encounter issues such as in estimation (due to complex likelihood functions) and in forecasting (since volatility drive forecast errors to explode). While volatility can be generalized to any of these models, ARCH(1) was used in Section 3. The algorithm presented in Section 3 can be modified minimally to consider a more general volatility model.

In analyzing multiple time series data, [9] developed a test for existence of dynamic orthogonal components. If dynamic orthogonal components indeed exist, [9] suggests that a univariate analysis of each time series will suffice. This is not the case in the event of a contagion. [10] observed that contagion in global stock markets are usually triggered by local crashes leading to region crashes.  [10] also noted that contagion is also affected by past occurrences of similar crashes, i.e., contagion is mathematically represented by conditional heteroscedasticity summarized in a typical volatility model.

While [11] noted that only the market itself contributes to volatility clustering in a global sense, locally, the cluster itself can contribute to volatility clustering effect, e.g., efficiency-inducing policies for a sector. There have been several works that further understand volatility clustering. [12] observed that clustered volatility is driven when funds are allowed to borrow from a bank, i.e., allowed to purchase more assets than their wealth would permit. Clustered volatility results from bank policies that forces values-at-risk of funds to behave similarly.

Using vine copula model to explain a dynamic behavior of multiple time series, [13] noted that the model accounts for volatility clustering, further highlighted that these models are very useful in value-at-risk forecasting (fewer capital requirements) since it produces smoother and more accurate forecasts.

[14] also used a new distance measure to cluster financial time series based on a variance ratio test statistic. The method aggregates time series according to autocorrelations, [14] observed that it discriminates stock markets reasonably well according to size and level of development.

## 3. Test for Volatility in Multiple Time Series

Given multiple time series, i.e., there are N time series each with T observations, [4] considered the following model:

$$Y_{i,t} = \phi Y_{i,t-1} + \lambda_i + u_{i,t},\ \ \lambda_i \sim \left(\mu_i, \sigma_{\lambda_i}^2\right)\ u_{i,t} \sim (0, \sigma_u^2) \tag{1}$$

for $i = 1,2, \dots, N$ and $t = 1,2, \dots, T$. Suppose the N time series is grouped into the m clusters, each with $n_j$ elements, $N = n_1 + \cdots + n_m$. The formulation in Model (1) do not consider relationship between $N$ and $T$, i.e., it is possible for $N < T, N > T,$ or $N = T$. Model (1) is modified to account for clustered conditional heteroscedasticity in the error term $u_{i,t}$ as follows:

$$Y_{i,t} = \phi Y_{i,t-1} + \lambda_i + u_{i,t},\ \ \lambda_i \sim \left(\mu_i, \sigma_{\lambda_i}^2\right),\ \ u_{i,t} = v_t \sigma_{kt} \tag{2}$$

where $\sigma_{kt}^2$ accounts for conditional heteroscedasticity present in cluster $k$, $k = 1, \dots, m$ and $v_t$ is a white noise process. This implies that all time series within each cluster exhibit similar volatility behavior, and volatility models may possibly vary across different clusters.

Suppose that volatility model for each cluster is ARCH (1), and that all time series within the cluster $(n_k)$ share similar parameters. The time series in other clusters may assume different parameters. There is a need to estimate the parameters shared by all time series within the cluster. On the other hand, totality of the series shares the same parameter $\phi$ in the dynamic model, but with varying random effects ($\lambda_i$) for every time series.

## Estimation Phase

The model is estimated in an iterative algorithm based on the backfitting framework in the algorithm below. Initialize $\hat{\lambda}_i$ by ignoring autoregressive and error terms from Equation (2). Other parameters are initialized by refitting of residuals in a backfitting algorithm. For the $b^{th}$ iteration:

1. Given recent estimates of the parameters, compute residuals from equation (2) except for the random effects, which is estimated from the residuals using the BLUP method, i.e., $\hat{\lambda}_i^{(b)} = \hat{\mu}_i$ since $E(\lambda_i) = \mu_i$.
2. Compute new residuals: $r_{it}^{*(b)} = Y_{it} - \hat{\lambda}_i^{(b)}$.

   Note: Rescaling of residuals $r_{it}^{*(b)}$ by the estimated volatility component $\hat{\sigma}_{it}^2$ is not necessary since the backfitting algorithm is fairly optimal with additivity of the model, see for example [15].
3. Estimate $\phi$ by $\hat{\phi}^{BS(b)}$ from the following bootstrap method sub steps:
   a. For each of the N time series of residuals $r_{it}^{*(b)}$, estimate $\phi$ as the autoregressive parameter and intercept of the residuals using conditional least squares to obtain $\hat{\phi}_i$.
   b. Resample from $\hat{\phi}_i, i = 1, \ldots, N$, to obtain $\hat{\phi}^{BS(b)}$ (simple random sample with replacement of size N, for R replicates). This is an ordinary bootstrap since each time series $i$ provided one estimate for the autoregressive parameter ($\hat{\phi}_i$).

4. Compute two forms of new residuals:

$$r_{it}^{**} = Y_{it} - \hat{\phi}^{BS(b)} Y_{i,t-1} \text{ and } r_{it}^{***} = Y_{it} - \hat{\lambda}_i^{(b)} - \hat{\phi}^{BS(b)} Y_{i,t-1}.$$

   Note: $r_{it}^{**}$'s will be used to estimate $\phi$ while $r_{it}^{***}$'s will be used to estimate the volatility model. Note that the bootstrap intercept is also subtracted from the $Y_{it}$'s in the $r_{it}^{***}$'s. Presence of volatility in the model affects the level of the residuals, and by subtracting the bootstrap intercept, stabilization in the levels of the random component is achieved.

5. For each time point $(t)$, define the square of $r_{it}^{***}$ as $\hat{u}_{it}^2$. Note that $r_{it}^{***}$ is a random sample of size 1. Thus, $\hat{\sigma}_{it}^2 = \hat{u}_{it}^2$ is an unbiased estimator of $\sigma_{it}^2$. Estimate the variance model, e.g., $(\sigma_{it}^2) = \alpha_{k,0} + \alpha_{k,1} u_{it-1}^2$ (for ARCH(1)) using $(\hat{u}_{it}^2, \hat{u}_{it-1}^2)$ thru OLS to obtain $\hat{\alpha}_{0t}^{(b)}$ and $\hat{\alpha}_{1t}^{(b)}$.
6. Estimate $\alpha_{k,0}$ by $\hat{\alpha}_{k,0}^{(b)}$ (the mean of $\hat{\alpha}_{0j}^{(b)}, j = 1, \dots, n_k$) and $\alpha_{k,1}$ by $\hat{\alpha}_{k,1}^{(b)}$ (the mean of $\hat{\alpha}_{1j}^{(0)}$ $j = 1, \dots, n_k$). $\hat{\alpha}_{0j}^{(0)}$ and $\hat{\alpha}_{1j}^{(0)}$ are estimates of ARCH(1) parameters among time series in cluster $k$, where $k = 1, 2, \dots, m$.

   Note: This implies that different ARCH (1) parameters are estimated for each cluster.

   Iterate from Step 1 estimating the random effects where $Y_{it}$ is replaced by $r_{it}^{**}$ and using $r_{it}^{*}$ as the adjusted time series data until the convergence, e.g., when parameter changes in-between iteration by less than the tolerance level $\varepsilon$.

## Testing for Volatility

Given parameter estimates from the Estimation Phase,

1. Reconstruct variance components for each resample thru $(\hat{\sigma}_{it}^2) = \hat{\alpha}_{k,0} + \hat{\alpha}_{k,1}\hat{u}_{it-1}^2$.
2. Generate $u_{it}^*$ from $N(0, \hat{\sigma}_{it}^2)$.
3. Compute replicates of $Y_{it}$ as $Y_{it}^* = \hat{\phi}Y_{it-1}^* + \hat{\lambda}_i + u_{it}^*$
4. Estimate parameters from each replicate of the data using the Estimation Algorithm above.

Multiple clusters are tested simultaneously. To control the familywise error rate (FWER), size $\alpha$ of the test is adjusted to $\alpha/m$ (Bonferroni correction) where $m$ is the number of clusters, see for example, [16]. Given the bootstrap replicates, $\left(\frac{\alpha}{2m}\right)^{th}$ and $\left(1 - \frac{\alpha}{2m}\right)^{th}$ percentiles of $\hat{\alpha}_{k,1}$s is computed and are used to test the significance of the parameter estimate for each cluster. Non-inclusion of zero in the interval provides enough empirical evidence against the null hypothesis (i.e., no significant volatility) while inclusion of zero indicates no evidence against the null hypothesis. For the variance model, $\alpha_{k,1} = 0$ indicates no volatility (assuming ARCH (1) model). Hence, the test is equivalent to the null which is absence of volatility of specific model, e.g., ARCH (1) again the alternative that volatility of specific model exists.

The method discussed above assumes that clusters are identified. Existence of clusters (number of clusters and membership of time series to a cluster) can be postulated by the analyst, e.g., stocks that are more likely involved in a possible contagion. Alternatively, number and cluster membership can be determined statistically through time series clustering, see for example [19].

# 4. Simulation Study

We designed a simulation study to investigate the computational optimality of the test. Some conditions about the data generating process are controlled, and this includes: number of time series (N=50), length of each time series (T=50); autoregressive parameter ($\phi$=0.6, 0.95 to represent stationary and near nonstationary time series, respectively); mean of random effect ($\mu_i = 0$); constant standard deviation of random effect across all time series; number of clusters (1 or 5, absence or presence of clustering, respectively); ARCH parameters [($\alpha_{k0} = 1, \alpha_{k1} = 1$)-presence of volatility, ($\alpha_{k0} = 1, \alpha_{k1} = 0$)-absence of volatility]; and when there are 5 clusters, 1 or 3 of the clusters are set to exhibit an ARCH(1) type of volatility. In all cases, level of significance is set at ($\alpha = 0.05$).

The data was simulated with Equation (2) as the data generating process. Random variables are first generated from the corresponding distribution. The white noise process $v_t$ was generated from the standard normal distribution. After initialization of the time series, repetitive substitution of previous values, assumed parameters, current and past values of random components to Equation (2) is done until 2T time points are generated. The first half of the simulated time series are dropped as this might have been influenced by initial values.

The nonparametric test is compared to a parametric test based on ARCH (1) model where each time series is treated in a univariate context. The parametric test for volatility is based on the likelihood ratio test, see for example [20]. The goal of the comparison is to assess whether knowledge of clustering can contribute in detecting group volatility. Power and size comparisons between parametric and nonparametric tests for various scenarios are summarized in Table 1.

**Table 1. Simulation Results for Scenarios without Misclassified Time Series in a Cluster**

| Scenario | Autoregressive Parameter ($\phi$) | Power of the Test | | Size of the Test | |
|---|---|---|---|---|---|
| | | Nonparametric | Parametric | Nonparametric | Parametric |
| Single Cluster | 0.6 | 1.0000 | 0.3762 | 0.0117 | 0.0224 |
| Single Cluster | 0.95 | 0.9081 | 0.2193 | 0.0000 | 0.0681 |
| 5 Clusters, Only 1 Cluster with Volatility | 0.6 | 0.6250 | 0.4110 | 0.0078 | 0.0236 |
| 5 Clusters, Only 1 Cluster with Volatility | 0.95 | 0.2711 | 0.2410 | 0.0042 | 0.0648 |
| 5 Clusters, With 3 Clusters with Volatility | 0.6 | 0.5854 | 0.3852 | 0.0061 | 0.0213 |
| 5 Clusters, With 3 Clusters with Volatility | 0.95 | 0.1383 | 0.2288 | 0.0000 | 0.0600 |

## Single Cluster, No Volatility

If all time series forms a single cluster, the nonparametric test is correctly-sized regardless on whether the time series are stationary (in mean) or nearly non-stationary. The parametric test is also correctly-sized when the time series is stationary in mean. However, size of the parametric test is distorted when the time series approaches nonstationarity in mean. This is not the case in the nonparametric test since all replicates under near nonstationarity failed to reject the null hypothesis of no volatility.

## Single Cluster, Volatility (ARCH) is Present

ARCH-type volatility model is induced to the simulated time series in cases where there is only a single cluster. The nonparametric test that consider all time series to provide evidence against the null hypothesis of no volatility yield very high power compared to the parametric counterpart that considers each time series individually, regardless of the state of stationarity in mean. In cases where the time series are stationary in mean, the nonparametric test was able to provide evidence against the null hypothesis for all replicates of the simulated data, while very low power was observed in the parametric test. As the time series approaches nonstationarity, both the nonparametric and parametric tests suffer a decline in power, but the decline in power of the parametric test is much larger than the decline in power of the nonparametric test (still exhibiting a reasonable power).

### 5 Clusters, Volatility (ARCH) is Present in 1 Cluster

Assuming 5 clusters, without inducing volatility in the simulated time series, both parametric and nonparametric test are correctly-sized. However, when the time series approaches nonstationarity, the parametric test already suffers from size distortion since the procedure relies heavily on the stationarity in mean assumption. This is not the case for the nonparametric test that is still correctly-sized even if the time series approaches near-nonstationarity. When volatility is induced in simulated time series in one cluster (time series in four other clusters do not contain volatility), the nonparametric test exhibit over 20% advantage in power compared to the parametric test in time series that are stationary in mean. When the time series approaches nonstationarity, both the parametric and nonparametric tests have lower power, the nonparametric test though still have relative advantage over the parametric test.

### 5 Clusters, Volatility (ARCH) is Present in 3 of the Clusters

Both the parametric and nonparametric tests are consistently correctly-sized when all the time series in 5 clusters exhibit stationarity in mean. The parametric test however, exhibit distortion in size when the time series in all clusters approaches nonstationarity in mean, this is not the case for the nonparametric test which is still correctly sized even when the time series approaches nonstationarity. As volatility is induced in three of the five clusters, the nonparametric test still has over 20% advantage in terms of power over the parametric test. Power of both parametric and nonparametric tests suffer as the time series across all clusters approaches nonstationarity.

## Misclassified Time Series

To verify robustness of the test to possible misclassification of time series into a cluster, a cluster of 50 time series with volatility is deliberately contaminated with some time series that does not exhibit volatility. Furthermore, similar cluster of 50 time series without volatility contaminated with some time series that actually exhibit volatility.

With 50 time series simulated to exhibit volatility, one time series (2%) or five (10%) time series that does not exhibit volatility were included. Provided that the time series are stationary (autoregressive parameter of 0.60), the test is able to identify volatility for all replicates. Relatively lower power (80%) is obtained when autoregressive parameter is 0.95.

The test is still able to detect even with only one (2%) or five (10%) time series with volatility are induced in a cluster of 50 time series. The chance of detecting volatility increases with more time series that actually exhibit volatilities in a cluster. Thus, regardless of the actual number of time series that exhibits volatility, the test is capable of detection of such. See Table 2 for details.

**Table 2. Simulation Results for Scenarios with Misclassified Time Series in a Cluster**

| **Scenario** | **Autoregressive Parameter ($\boldsymbol{\phi}$)** | $\boldsymbol{P}$(Rejecting $\boldsymbol{H_0}$) |
|---|---|---|
| No Volatility (2% with Volatility) | 0.60 | 0.1162 |
| No Volatility (10% with Volatility) | 0.60 | 0.4731 |
| No Volatility (2% with Volatility) | 0.95 | 0.2062 |
| No Volatility (10% with Volatility) | 0.95 | 0.2513 |
| With Volatility (2% No Volatility) | 0.60 | 1.0000 |
| With Volatility (10% No Volatility) | 0.60 | 1.0000 |
| With Volatility (2% No Volatility) | 0.95 | 0.8077 |
| With Volatility (10% No Volatility) | 0.95 | 0.7913 |

# 5. Application in Stock Market Price Indices

Contagion is a common event in stock markets usually resulting from interdependence among securities and among stock brokers. Volatility is another stylized fact among indicators that characterizes behavior of the market, often monitored at very high frequencies by various stakeholders. [17] noted that there is evidence of contagion from the US stock market to Japan, United Kingdon, France, Germany, Hong Kong, and Canada. They further noted that contagion is not just a crisis-specific event, but is present in the market all the time. [18] further observed that during the major crisis in European equity markets, contagion effects generated short-term shocks, also noted that there is evidence that the most recent US subprime crisis is brought about by contagion effect. These short-term shocks can easily drive volatility of key market indicators like prices.

We used prices of stocks traded in the European and US markets to investigate presence of volatility associated with contagion effect. Regional contagion can cause volatility among stock market prices in the region. In understanding the dynamic behavior of stock prices, time series data of prices of 30 stocks are postulated to cluster into European (19 stocks) and US (11 stocks) regions. Daily prices during 2011-2016 period are used in the analysis.

## Original Time Series Data

Six of nineteen European stocks are plotted in Figure 1, while six of the eleven stocks in the US market are plotted in Figure 2. While there are some periods where volatility seems to exists, this can potentially be masked by overall nonstationarity. From Table 3, The original time series data both from the European and US markets exhibit nonstationarity, most of the estimated autoregressive parameters are 0.99 or 0.98, smallest value was in a stock in the US market where autoregressive parameter is 0.937.With the original time series data, nonstationarity in mean is dominating, so that the parametric test for volatility failed to reject the null hypothesis of no volatility for all time series, see Table 3 for details.

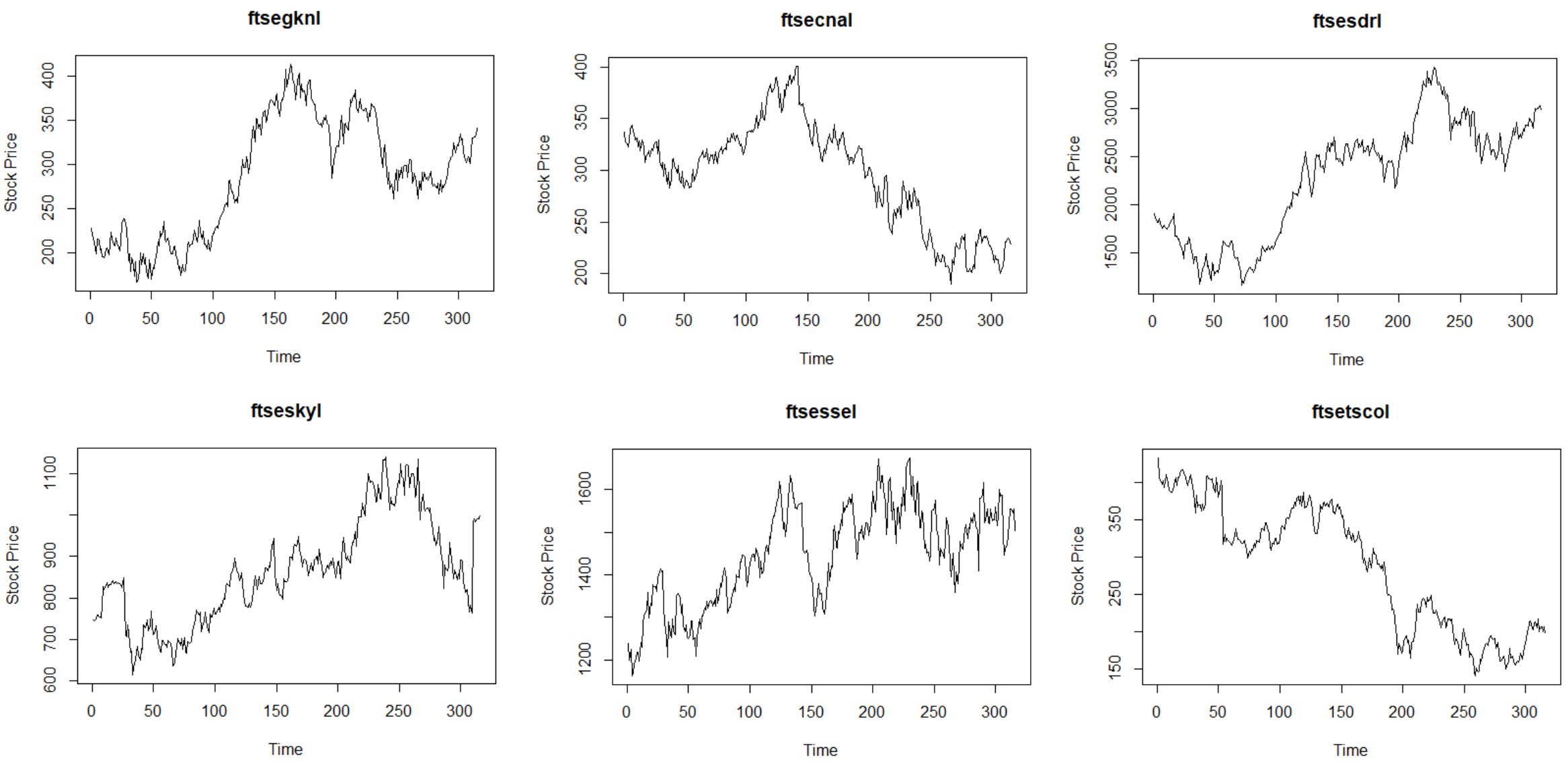

**Figure 1. Time Plot of Some European Stock Prices**

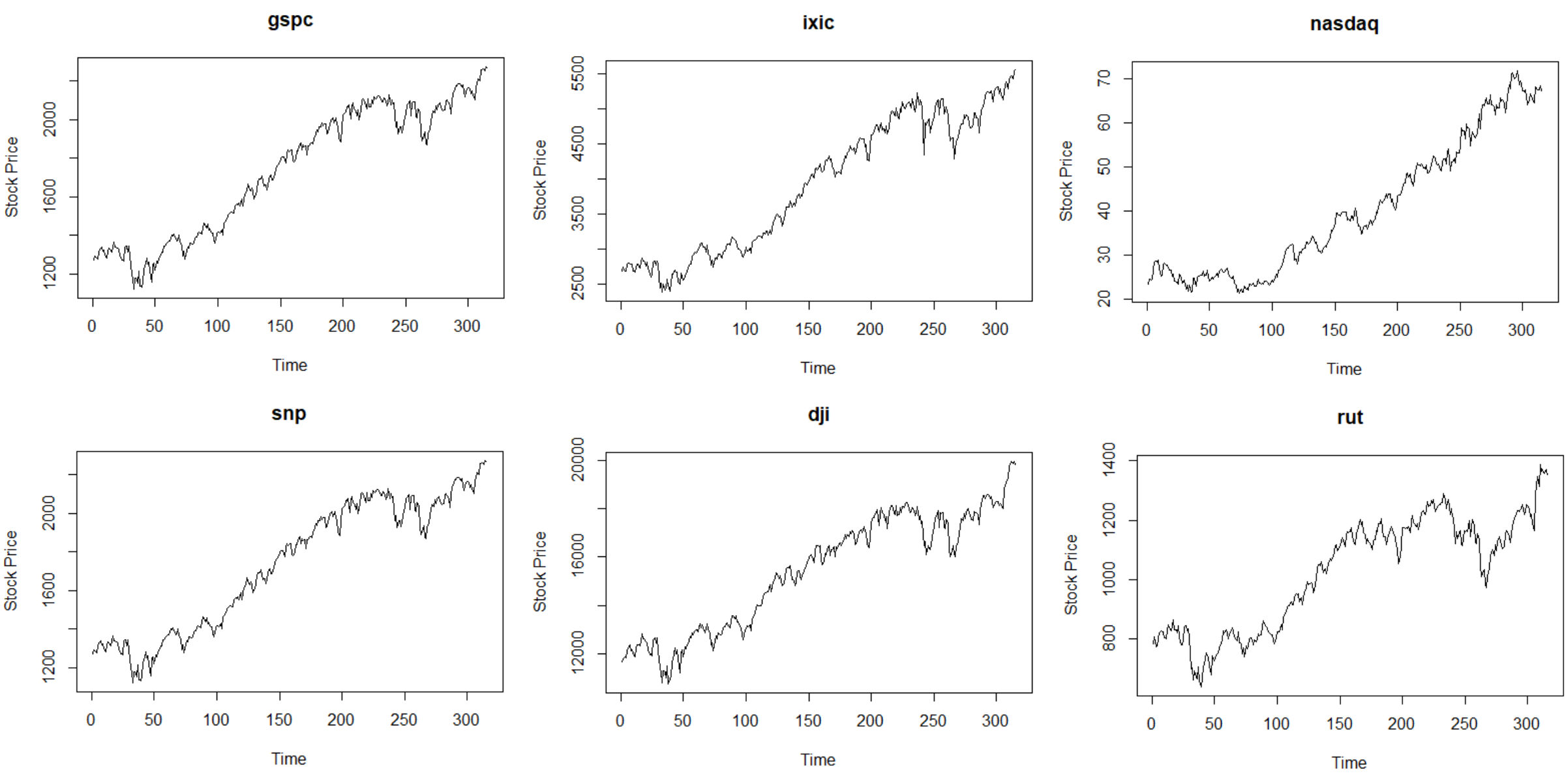

**Figure 2. Time Plot of Some US Stock Prices**

**Table 3. Univariate Analysis of 30 Time Series Data**

| Stocks | Cluster | AR (1) Estimate | p-value of parametric test assuming ARCH (1) |
|---|---|---|---|
| gdaxi | Europe | 0.991521 | 0.452747 |
| ftseanthl | Europe | nonstationary | --- |
| ftseantoi | Europe | 0.987608 | 0.211045 |
| ftsebal | Europe | 0.992843 | 0.755044 |
| ftsebatsl | Europe | 0.993605 | 0.895994 |
| ftsegknl | Europe | 0.985435 | 0.148366 |
| ftsecnal | Europe | 0.986601 | 0.715972 |
| ftsepfgl | Europe | 0.996475 | 0.247612 |
| ftsepnsl | Europe | 0.995587 | 0.965042 |
| ftseprul | Europe | 0.993537 | 0.639695 |
| ftserbl | Europe | 0.996969 | 0.729532 |
| ftserrl | Europe | 0.987278 | 0.97215 |
| ftsesdrl | Europe | 0.990858 | 0.052396 |
| ftseshpl | Europe | 0.992136 | 0.189234 |
| ftseskyl | Europe | 0.973691 | 0.838058 |
| ftsessel | Europe | 0.951328 | 0.303283 |
| ftsestjl | Europe | 0.996451 | 0.767607 |
| ftsetscol | Europe | 0.994495 | 0.549249 |
| ftsevodl | Europe | 0.989421 | 0.937241 |
| gspc | US | 0.997792 | 0.66884 |
| ixic | US | 0.997112 | 1.19E-07 |
| nasdaq | US | 0.998013 | 0.284526 |
| nya | US | 0.990934 | 0.149703 |
| rut | US | 0.993939 | 0.508075 |
| snp | US | 0.997792 | 0.66884 |
| ta | US | 0.969655 | 0.829248 |
| tsx | US | 0.956174 | 0.17366 |
| xax | US | 0.937966 | 0.912286 |
| bvsp | US | 0.972485 | 0.082685 |

Using the estimation procedure for clustered time series data described in Section 3, parameters of the mean and variance models are estimated per cluster and presented in Table 4. In the multiple time series framework, we assumed similar model for the mean of the time series. The common autoregressive parameter is estimated at 0.9863, which is within the values of estimated autoregressive parameters (univariate) for the individual time series in Table 3.

**Table 4. Estimate of the Common Autoregressive Parameter and the ARCH (1) Parameters per Region**

| **Autoregressive Parameter ($\widehat{\phi}$)** | **Volatility Slope Coefficient ($\widehat{\alpha}_{k,1}$)** | |
|---|---|---|
| | European | US |
| 0.9863 | 2.980<br>(0.4245,3.4143)* | -0.021<br>(-0.1476,0.1731)* |

**Bonferroni corrected 95% Confidence Interval*

The Bonferroni corrected CI for the European market do not include 0, indicating that as a cluster, the European market exhibits volatility. Note that consistent with results of simulation studies, the parametric test failed to provide empirical evidence on the existence of volatility, while the nonparametric test was able to recognize empirical evidence of joint volatility (possibly caused by contagion) among the stocks in the European market. The Bonferroni corrected CI for the US market includes 0, hence, even the nonparametric test failed to recognize empirical evidence of the existence of group volatility among the stocks in the US market. Power of the nonparametric test diminish when the time series are nearly nonstationary.

## First Differenced Time Series

The parametric test for volatility suffers from size distortion when the time series approaches nonstationarity, which is not the case in the nonparametric test. Also, power is reduced even in the nonparametric test as the time series approaches nonstationarity, but with greater reduction in power for the parametric test. First differences of the time series are obtained to mitigate presence of nonstationarity. Time plots of six stocks in the European market are given in Figure 3 and the time plots of six stocks in the US market are given in Figure 4. Both clusters now exhibit stationary behavior and volatility has become more visually evident.

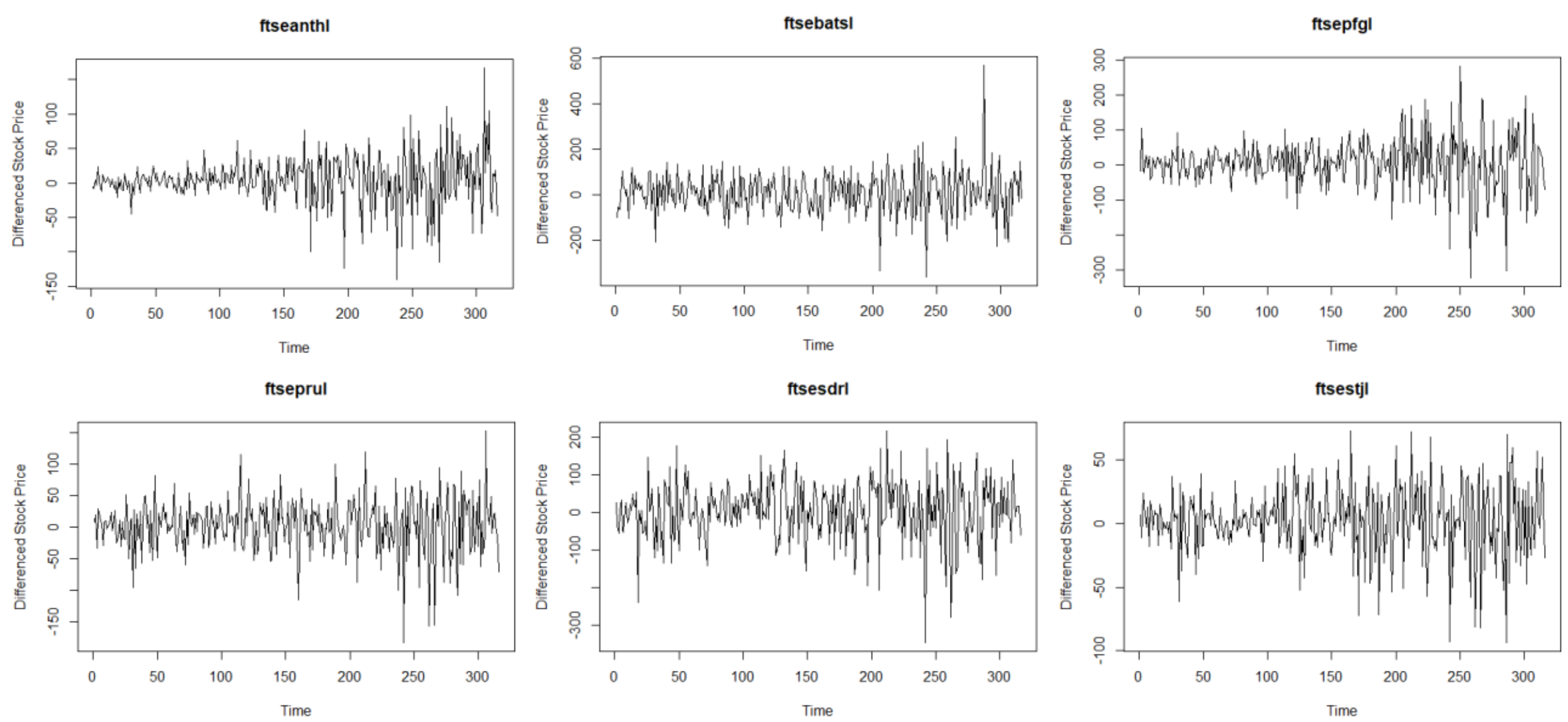

**Figure 3 Time Plot of Some First Differenced European Stock Prices**

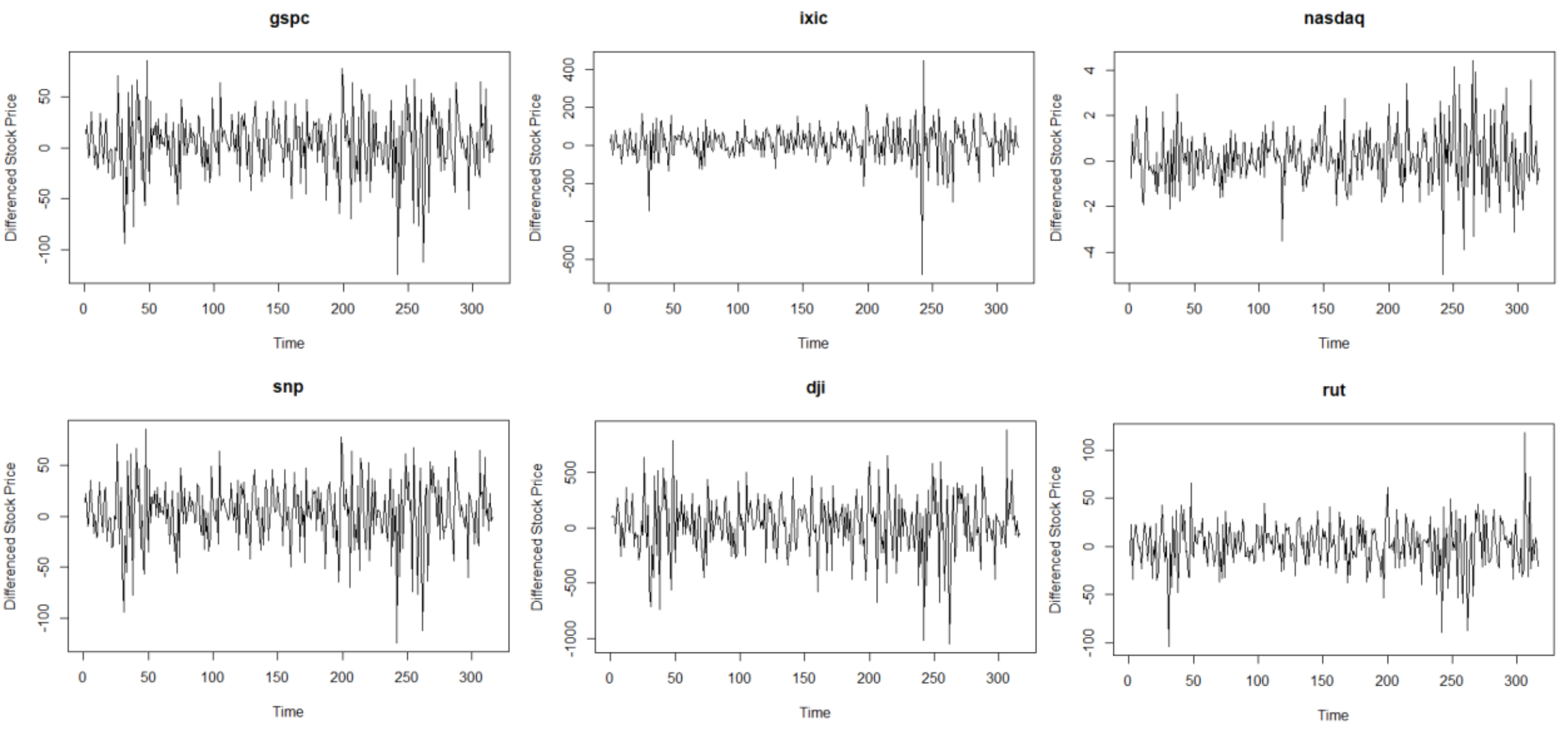

**Figure 4 Time Plot of Some First Differenced US Stock Prices**

Univariate analysis was done with the individual (first-differenced) time series, estimates and results of parametric tests for volatility are summarized in Table 5. All first differenced time series are now stationary. In fact, many of the time series are actually random walk since no dependence structure is evident from the first differenced time series. Only four stocks in the European market and two stocks in the US market still exhibit dependencies after first differencing. The parametric test for volatility identifies only one time series in the European and one in the US market to exhibit volatility.

**Table 5. Univariate Analysis of 30 Time Series Data (First Differenced)**

| Stocks | Cluster | AR (1) Coefficient | | p-value of parametric test assuming ARCH (1) |
|---|---|---|---|---|
| | | Estimate | p-value | |
| gdaxi | Europe | -0.07667 | 0.171013 | 0.484642 |
| ftseanthl | Europe | -0.15284 | 0.006077 | 0.962022 |
| ftseantoi | Europe | -0.10686 | 0.059935 | 0.363237 |
| ftsebal | Europe | -0.10563 | 0.059022 | 0.625738 |
| ftsebatsl | Europe | -0.01579 | 0.779105 | 0.965093 |
| ftsegknl | Europe | -0.17087 | 0.002034 | 0.459401 |
| ftsecnal | Europe | -0.07678 | 0.170875 | 0.726868 |
| ftsepfgl | Europe | -0.04521 | 0.421263 | 0.255153 |
| ftsepnsl | Europe | -0.11653 | 0.037212 | 0.821268 |
| ftseprul | Europe | -0.0874 | 0.120066 | 0.632106 |
| ftserbl | Europe | -0.08967 | 0.109265 | 0.598744 |
| ftserrl | Europe | -0.03911 | 0.486438 | 0.980128 |
| ftsesdrl | Europe | -0.02777 | 0.621233 | 0.052873 |
| ftseshpl | Europe | -0.03354 | 0.552309 | 0.24739 |
| ftseskyl | Europe | -0.08359 | 0.135394 | 0.747185 |
| ftsessel | Europe | -0.1061 | 0.058469 | 0.275747 |
| ftsestjl | Europe | -0.1211 | 0.030172 | 0.877905 |
| ftsetscol | Europe | -0.05697 | 0.31021 | 0.769745 |
| ftsevodl | Europe | -0.00777 | 0.890174 | 0.936011 |
| gspc | US | -0.0936 | 0.094255 | 0.604614 |
| ixic | US | -0.17454 | 0.0016 | 0.008302 |
| nasdaq | US | -0.14845 | 0.007587 | 0.975516 |
| nya | US | -0.10032 | 0.072651 | 0.401013 |
| rut | US | -0.04546 | 0.418536 | 0.479858 |
| snp | US | -0.0936 | 0.094255 | 0.604614 |
| ta | US | 0.032897 | 0.55791 | 0.899131 |
| tsx | US | -0.08347 | 0.136433 | 0.377397 |
| xax | US | -0.08884 | 0.113139 | 0.608238 |
| bvsp | US | -0.01567 | 0.780449 | 0.110908 |
| dji | US | -0.09195 | 0.100237 | 0.360457 |

We also used the estimation procedure for clustered data described in Section 3 for the first differenced time series. Parameters of the mean and variance models are estimated per cluster and presented in Table 6. From the multiple time series assumption, the common autoregressive parameter is estimated to be -0.0809, within the range of values of the autoregressive coefficients from the univariate analysis in Table 5.

**Table 6. Estimate of the Common Autoregressive Parameter and the ARCH (1) Parameters per Region (First Differenced)**

| **Autoregressive Parameter ($\hat{\phi}$)** | **Volatility Slope Coefficient ($\hat{\alpha}_{k,1}$)** | |
|---|---|---|
| | European | US |
| -0.0809 | 33.85<br>(1.2952, 27.4435)* | 1.02<br>(0.8197, 1.5156)* |

**Bonferroni corrected 95% Confidence Interval*

From Table 6, the Bonferroni corrected CI for the European market do not include 0, indicating that as a cluster, the European market exhibits volatility. Similar is true for the US market, the Bonferroni corrected CI also precludes zero, indicating presence of volatility among the clustered time series. Recall that the simulation study indicates higher power for the nonparametric test when the individual time series are stationary. While the parametric test for volatility in Table 5 identifies only one time series to exhibit volatility, the nonparametric test in Table 6 provides empirical evidence that clustered volatility is present in both the European and US markets.

## 6. Conclusions

Given clustered time series data, a nonparametric test for volatility is proposed, this accounts for the possible contagion effect among time series in the same cluster. The simulation study illustrate that the test is correctly-sized even when the multiple time series approaches nonstationarity. The test is powerful if volatility is contained in fewer clusters only, a resemblance of localized contagion effect. As contagion causing volatility become global in nature, i.e., as more clusters are affected by volatility, even the nonparametric test exhibits low power. Note however that widespread volatility, i.e., practically all time series manifest volatility behavior, is also the case where volatility often becomes more obvious even visually. The nonparametric test offers a method of testing volatility in multiple time series that exhibit clustering, and that volatility spillover is contained only in few clusters. In the presence of contagion, whether local or global, the nonparametric test can benefit from the simultaneous evidence that all time series can provide against absence of volatility. A clear understanding of presence of volatility will facilitate identification and estimation of models that can generate reliable forecast of indicators involved, hence, better risk management in sectors that manifest such volatile behavior like the financial markets.